\documentclass{webofc}

\usepackage[varg]{txfonts}   
\usepackage{hyperref}
\usepackage{url}
\hypersetup{colorlinks=true,citecolor=blue,urlcolor=blue,linkcolor=blue}
\begin{document}
\title{Isospin-symmetry breaking in kaon production: the role of charge-symmetry}
%
%

\author{\firstname{Francesco} \lastname{Giacosa}\inst{1,2}\fnsep\thanks{\email{fgiacosa@ujk.edu.pl}} 
}

\institute{Institute of Physics, Jan Kochanowski University, ul. Uniwersytecka 7, 25-406, Kielce, Poland
\and
          Institute for Theoretical Physics, J. W. Goethe University,
Max-von-Laue-Straße. 1, 60438 Frankfurt am Main, Germany
          }

\abstract{An excess of charged versus neutral kaons has been recently
reported by the NA61/SHINE collaboration.  Similar excesses were
also present in previous experiments, even if with larger errors. Models for
hadron productions in heavy ion collisions systematically underestimate the
measured charge-to-neutral kaon ratio. In this report, we comment on the
role of charge-symmetry as a specific isospin transformation. In particular,
we show that the ensemble of the initial colliding nuclei being
charge-symmetric is sufficient for producing an equal number of charged and neutral kaons
in the isospin-symmetry exact
limit. 
}
\maketitle
\section{Introduction}
\label{intro}
A surprisingly large excess
of charged over neutral kaons has been reported in Ar-Sc scattering by the NA61/SHINE experiment at CERN \cite{na61}.
As argued in the companion theoretical paper in Ref. \cite{brylinski}, two well
established models for heavy ion collisions, the hadron resonance gas (HRG)
and the Ultrarelativistic Quantum Molecular Dynamics (UrQMD) approaches \cite{bleicher,vovchenko}, are incapable of reproducing the experimental data. 
The HRG and
UrQMD theoretical results are systematically smaller than the experimental ones,
resulting in a theory-to-experiment mismatch of almost $5\sigma $
\cite{brylinski}. 

The charged-to-neutral kaon ratio is interesting since it quantifies isospin
breaking. If an initial state with two nuclei  containing
an equal number of protons and neutrons is considered ($Q/B=1/2$), isospin
symmetry `implies' an equal number of charged and neutral kaons.
In this work we concentrate on this point. To this end, we make use of charge symmetry, which is a specific isospin symmetry that interchanges a member of a
given isospin multiplet with $(I,I_{z})$ with the member $(I,-I_{z}),$ as
for instance: $p\leftrightarrows n,$ $K^{+}\leftrightarrows K^{0},$ $\bar{K}%
^{0}\leftrightarrows K^{-},$ $\pi ^{+}\leftrightarrows \pi ^{-}$ etc.
In particular, we show that the initial ensemble of scattering nuclei being
charge-symmetry invariant leads to the prediction that the total charged and neutral
kaon multiplicities are equal. 

\section{Charge-symmetry for the initial ensemble}
\label{sec-1}
The isospin operator $\hat{\mathbf{I}}=(\hat{I}_{x},\hat{I}_{y},\hat{I}_{z})$
enters into the $SU(2)$ internal rotation described by the unitary operator $%
\hat{U}_{I}=e^{i\theta _{k}\hat{I}_{k}}$ (with $k=1,2,3$): in the fundamental
representation, $\hat{I}_{k}=\sigma _{k}/2$ where $\sigma _{k}$ are the
Pauli matrices, according to which the light quark `isodoublet' transforms
as $(u,d)^{t}\rightarrow U_{I}(u,d)^{t}$. As a consequence, the quark $u$
has eigenvalue $I_{z}=1/2$, the quark $d$ the eigenvalue $I_{z}=-1/2$ and
both quarks carry the eigenvalue $I(I+1)=3/2$ for the operator $\hat{\mathbf{%
I}}^{2},$ thus $I=1/2$. The antiquark doublet $(-\bar{d},\bar{u})^{t}$
transforms just as the quark one.

Kaons are grouped into the isodoublet $(I=1/2)$ states $(K^{+}\equiv u\bar{s}%
,K^{0}\equiv d\bar{s})$ and $(-\bar{K^{0}}\equiv -s\bar{d},K^{-}\equiv s\bar{%
u})$ that correspond naturally to the underlying isospin doublets $(u,d)$
and $(-\bar{d},\bar{u})$. Namely, they transform as (anti)quarks
because the $s$-quark is left invariant by isospin. 

Isospin is well conserved in low-energy QCD processes: the particle masses
well reflect isospin symmetry \cite{pdg}, e.g. $\left( m_{\pi ^{+}}-m_{\pi ^{0}}\right)
/\left( m_{\pi ^{+}}+m_{\pi ^{0}}\right) \simeq 0.017$ and $\left(
m_{K^{+}}-m_{K^{0}}\right) /\left( m_{K^{+}}+m_{K^{0}}\right) \simeq -0.004$, and the treatment of pion-pion, pion-kaon, pion-nucleon, and
nucleon-nucleon scattering closely follows the predictions of isospin
symmetry~\cite{pennington,alarcon}, see also the low-energy
phenomenology within the extended linear sigma model \cite{dick} (for a recent review see Ref. \cite{elsmrev}).

A specific isospin transformation, called charge symmetry $\hat{C}_{I}$ (not
be confused with charge conjugation $\hat{C}$ that swaps particles in
antiparticles) is realized by choosing $\hat{C}_{I}=e^{i\pi \hat{I}_{y}}$
(thus for $\theta _{1,3}=0$ and $\theta _{2}=\pi $). For the (anti)quarks
and kaon doublets, $\hat{C}_{I}$ takes the matrix form 
\begin{equation}
\hat{C}_{I}=\left( 
\begin{array}{cc}
0 & 1 \\ 
-1 & 0%
\end{array}%
\right) \text{ ,}
\end{equation}%
thus interchanging $u\leftrightarrow d$ and $\bar{u}\leftrightarrow \bar{d}$, 
and conversely $K^{+}\leftrightarrow $ $K^{0}$ and $K^{-}\leftrightarrow $ $%
\bar{K}^{0}$.

Next, we move to heavy-ion collisions and the related production of
kaons. Let us consider, for simplicity, a certain reaction to the
production of a single $K^{+}$: $A+A\rightarrow K^{+}X$, where $X$ refers to other particles (but no kaons).
Let us first assume that the initial state has total isospin $I=0.$ This
case corresponds to both nuclei having $I=0$. 
Indeed, nuclei usually appear in the lowest possible isospin state \cite{lenzi}. The
corresponding matrix element is $A_{K^{+}X}^{(0)}=\left\langle K^{+}X|\hat{S}%
_{QCD}|AA,I=0\right\rangle ,$ where $\hat{S}_{QCD}$ refers to the $\hat{S}$%
-matrix of QCD.
Isospin symmetry implies that $\hat{U}_{I}^{\dagger }\hat{S}_{QCD}\hat{%
U}_{I}\simeq \hat{S}_{QCD}$. Considering the charge transformation $\hat{U}%
_{I}=\hat{C}_{I}$ we get: 
\begin{equation}
A_{K^{+}X}^{(0)}=\left\langle K^{+}X|\hat{S}_{QCD}|AA,I=0\right\rangle
 \simeq \left\langle K^{0}\tilde{X}|\hat{S}%
_{QCD}|AA,I=0\right\rangle =A_{K^{0}\tilde{X}}^{(0)}
\text{ ,}
\end{equation}%
where $\hat{C}_{I}\left\vert AA,I=0\right\rangle =\left\vert
AA,I=0\right\rangle $ and $\hat{C}_{I}\left\vert X\right\rangle
=\left\vert \tilde{X}\right\rangle $ have been used, the latter being the charge--symmetry rotated `rest'.
Above, $A_{K^{0}\tilde{X}}^{(0)}$ is the amplitude for the $K^{0}\tilde{X}$
formation. Thus, the (approximate) equality $A_{K^{+}X}^{(0)}\simeq A_{K^{0}%
\tilde{X}}^{(0)}$ implies that $A+A\rightarrow K^{0}\tilde{X}$ is equally
probable as $A+A\rightarrow K^{+}X$. 
Extending the reasoning to any reaction involving kaons, 
one find that the number of produced $K^{+}$ equals the number of produced $K^{0}$
($\langle K^{+}\rangle =\langle K^{0}\rangle $.). The same argument
applied to the isodoublet pair $K^{-}$ and $\bar{K}^{0}$ leads to $\langle
K^{-}\rangle =\langle \bar{K}^{0}\rangle$.

Next, we consider a different initial state. For simplicity, 
for total isospin $I=1$ one has
\begin{equation}
A_{K^{+}X}^{(1)}=\left\langle K^{+}X|\hat{S}_{QCD}|AA,I=1\right\rangle \simeq -\left\langle K^{0}\tilde{X}|\hat{S}%
_{QCD}|AA,I=1\right\rangle =-A_{K^{0}\tilde{X}}^{(1)}
\text{ ,}
\end{equation}%
with $\hat{C}_{I}\left\vert AA,I=1\right\rangle =-\left\vert
AA,I=1\right\rangle $. Clearly, the opposite sign does not change the final
outcome for the cross-sections: $\langle K^{+}\rangle =\langle K^{0}\rangle $
and, analogously, $\langle K^{-}\rangle =\langle \bar{K}^{0}\rangle $ hold
also in this case.\ This result can be easily generalized to any initial
state with definite fixed isospin $I=I_{0}$ with $\hat{C}_{I}\left\vert
AA,I=I_{0}\right\rangle =(-1)^{I_{0}}\left\vert AA,I=I_{0}\right\rangle .$

The arguments above do not, however, imply that $\langle K^{+}\rangle
=\langle K^{0}\rangle $ and $\langle K^{-}\rangle =\langle \bar{K}%
^{0}\rangle $ hold in the most general case. As an illustrative example, we 
consider the following superposed initial state:%
\begin{equation}
\left\vert AA\right\rangle =\alpha _{0}\left\vert AA,I=0\right\rangle
+\alpha _{1}\left\vert AA,I=1\right\rangle 
\text{ .}
\label{aa}
\end{equation}%
The amplitude for $K^{+}X$ production is given by%
\begin{equation}
A_{\left\vert AA\right\rangle \rightarrow K^{+}X}=\alpha
_{0}A_{K^{+}X}^{(0)}+\alpha _{1}A_{K^{+}X}^{(1)}\rightarrow \left\vert
A_{\left\vert AA\right\rangle \rightarrow K^{+}X}\right\vert ^{2}=\left\vert
\alpha _{0}A_{K^{+}X}^{(0)}+\alpha _{1}A_{K^{+}X}^{(1)}\right\vert ^{2}.
\end{equation}%
Using analogous steps as before:%
\begin{equation}
A_{\left\vert AA\right\rangle \rightarrow K^{0}\tilde{X}}=\alpha _{0}A_{K^{0}%
\tilde{X}}^{(0)}+\alpha _{1}A_{K^{0}\tilde{X}}^{(1)}=\alpha
_{0}A_{K^{+}X}^{(0)}-\alpha _{1}A_{K^{+}X}^{(1)}
\rightarrow
\left\vert A_{\left\vert AA\right\rangle \rightarrow K^{0}\tilde{X}%
}\right\vert ^{2}=\left\vert \alpha _{0}A_{K^{+}X}^{(0)}-\alpha
_{1}A_{K^{+}X}^{(1)}\right\vert ^{2}
\text{ .}
\end{equation}%
Because of the different signs $%
\left\vert A_{\left\vert AA\right\rangle \rightarrow K^{+}X}\right\vert
^{2}\neq \left\vert A_{\left\vert AA\right\rangle \rightarrow K^{0}\tilde{X}%
}\right\vert ^{2}$, implying that the production probabilities for $K^{+}$ and 
$K^{0}$ are different for the initial state $\left\vert
AA\right\rangle$ of Eq. \ref{aa}.

At this point, we assume that the initial ensemble is charge-symmetric. That
means that the state%
\begin{equation}
\left\vert AA\right\rangle _{C_{I}}=C_{I}\left\vert AA\right\rangle =\alpha
_{0}\left\vert AA,I=0\right\rangle -\alpha _{1}\left\vert
AA,I=1\right\rangle 
\end{equation}%
occurs as often as the $\left\vert AA\right\rangle $ state of Eq. (\ref{aa}). By repeating the
previous passages:
\begin{equation}
\left\vert A_{\left\vert AA\right\rangle _{C_{I}}\rightarrow
K^{+}X}\right\vert ^{2}=\left\vert \alpha _{0}A_{K^{+}X}^{(0)}-\alpha
_{1}A_{K^{+}X}^{(1)}\right\vert ^{2}\text{ , }\left\vert A_{\left\vert
AA\right\rangle _{C_{I}}\rightarrow K^{0}\tilde{X}}\right\vert
^{2}=\left\vert \alpha _{0}A_{K^{+}X}^{(0)}+\alpha
_{1}A_{K^{+}X}^{(1)}\right\vert ^{2}.
\end{equation}%
Since $\left\vert AA\right\rangle $ and $\left\vert AA\right\rangle _{C_{I}}$
are equally probable, the final $K^{+}X$ rate is proportional to: 
\begin{equation}
\frac{1}{2}\left\vert A_{\left\vert AA\right\rangle \rightarrow K^{+}X}\right\vert
^{2}+\frac{1}{2}\left\vert A_{\left\vert AA\right\rangle _{C_{I}}\rightarrow
K^{+}X}\right\vert ^{2}=\left\vert \alpha _{0}A_{K^{+}X}^{(0)}\right\vert
^{2}+\left\vert \alpha _{1}A_{K^{+}X}^{(1)}\right\vert ^{2}\text{ ,}
\end{equation}%
where the interference terms disappear and equals the $K^{0}\tilde{X}$
production rate: 
\begin{equation}
\frac{1}{2}\left\vert A_{\left\vert AA\right\rangle \rightarrow K^{0}\tilde{X}%
}\right\vert ^{2}+\frac{1}{2}\left\vert A_{\left\vert AA\right\rangle
_{C_{I}}\rightarrow K^{0}\tilde{X}}\right\vert ^{2}=\left\vert \alpha
_{0}A_{K^{+}X}^{(0)}\right\vert ^{2}+\left\vert \alpha
_{1}A_{K^{+}X}^{(1)}\right\vert ^{2}\text{ .}
\end{equation}%
Thus, upon averaging, the final state $K^{+}X$ appears
just as often as $K^{0}\tilde{X}$. Of course, this result applies if the
initial ensemble is charge-symmetric. This point is well upheld because 
 nuclei can be well described as
clusters of $\alpha $-particles \cite{otsuka}.

These arguments can be repeated for any reaction involving kaons. Upon averaging over the initial states of a charge-symmetric
ensemble guarantees that $\langle K^{+}\rangle =\langle K^{0}\rangle $ (and
similarly that $\langle K^{-}\rangle =\langle \bar{K}^{0}\rangle $). Then,
the charged and neutral multiplicities coincide:  
\begin{equation}
\langle K^{+}\rangle +\langle K^{-}\rangle =\langle K^{0}\rangle +\langle 
\bar{K}^{0}\rangle \text{ .}
\end{equation}
The neutral kaons are not directly measured, since the physical states are the $K_{L}$ and $K_{S}$ which, by neglecting the (very) small CP
violation, are given by:
$K_{S(L)} = (K^{0} \mp \bar{K}^{0})/\sqrt{2}$. 
The emerging number of $K_{S}$ and $K_{L}$ is given by:
$\langle K_{S}\rangle =\frac{1}{2}\langle K^{0}\rangle +\frac{1}{2}\langle 
\bar{K}^{0}\rangle \text{ }=\langle K_{L}\rangle \text{ .}
$
The final expected relation is $\langle K^{+}\rangle +\langle
K^{-}\rangle =2\langle K_{S}\rangle ,$ implying that the ratio
\begin{equation}
R_{K}=\frac{\langle K^{+}\rangle +\langle K^{-}\rangle }{2\langle
K_{S}\rangle }=1  \label{Rk}
\end{equation}%
for colliding nuclei in the isospin-symmetric limit. This is why in Refs. \cite{na61,brylinski}
the ratio $R_{K}$ is studied both experimentally and theoretically.

\section{Summary}
In this work, we have shown that a charge-symmetric initial ensemble implies an equal number of
charged and neutral kaons in the isospin-exact limit.
As a consequence, the ratio $R_{K}$ of Eq. (\ref{Rk}) is unity. Violations from
this prediction appear because isospin symmetry is not exact.
However, these deviations ought to be small, as it is confirmed in actual
models. Namely, only in specific cases, the small differences between the
masses of charged and neutral kaons can lead to sizable modifications. A
typical example is the decay of the $\phi (1024)$ meson into kaons, for
which $\phi (1024)\rightarrow K^{+}K^{-}$ is 1.4 times larger  than $\phi
(1024)\rightarrow K^{0}$ $\bar{K}^{0}$ . Yet, the
sum of these effects leads only to a few percent modification of the
relation $R_{K}=1.$ At high scattering energies ($\sqrt{s_{NN}}\gtrsim 15$
GeV), one finds that $R_{K}\simeq 1.05,$ which is sizably smaller than the
central experimental values ($R_{K}\simeq 1.18$) of Ref. \cite{na61}. 

In the actual realized experiments the ratio $Q/B\simeq 0.4<1/2.$ The
surplus of neutrons favors the production of neutral kaons. This
effect is small at high energies but becomes important at small ones ($%
\sqrt{s_{NN}}\lesssim 10$ GeV). The final
mismatch of theory and experiment is quantified in Ref. \cite{brylinski} finding an
overall $\sim 5\sigma $ effect. 
Future efforts to explain this  `kaon anomaly' are needed. On the theoretical
side, one may explore models that may contain strong isospin breaking \cite{rafelski} (a
possible role could be played by electromagnetic interaction). On the
experimental side, the realization of experiments with nuclei with $Q/B=1/2$ is highly desirable.

 \bigskip
 
 \textbf{Acknowledgments}
 The author thanks the co-authors of Ref. \cite{brylinski}, A. Rybicki, L. Turko, and S. Mrówczyński for very useful discussions.

\end{document}